\documentclass[twocolumn,english,PRA twocolumn]{revtex4}
\usepackage[T1]{fontenc}
\usepackage[latin9]{inputenc}
\usepackage{amsbsy}
\usepackage{amssymb}
\usepackage{graphicx}
\usepackage{esint}
\usepackage{epstopdf}
\makeatletter
\@ifundefined{textcolor}{}
{%
 \definecolor{BLACK}{gray}{0}
 \definecolor{WHITE}{gray}{1}
 \definecolor{RED}{rgb}{1,0,0}
 \definecolor{GREEN}{rgb}{0,1,0}
 \definecolor{BLUE}{rgb}{0,0,1}
 \definecolor{CYAN}{cmyk}{1,0,0,0}
 \definecolor{MAGENTA}{cmyk}{0,1,0,0}
 \definecolor{YELLOW}{cmyk}{0,0,1,0}
}

\makeatother

\usepackage{babel}
\begin{document}

\title{Generation of Massive Entanglement Through Adiabatic Quantum Phase
Transition in a spinor condensate}

\author{Z. Zhang, L.-M. Duan}

\affiliation{Department of Physics, University of Michigan, Ann Arbor,
Michigan 48109, USA \\
and Center for Quantum Information, IIIS, Tsinghua University, Beijing 100084, China}

\begin{abstract}
We propose a method to generate massive entanglement in a spinor Bose-Einstein
condensate from an initial product state through adiabatic sweep of
magnetic field across a quantum phase transition induced by competition
between the spin-dependent collision interaction and the quadratic
Zeeman effect. The generated many-body entanglement is characterized
by the experimentally measurable entanglement depth in the proximity
of the Dicke state. We show that the scheme is robust to practical noise
and experimental imperfection and under realistic conditions it is
possible to generate genuine entanglement for hundreds of atoms.
\end{abstract}

\maketitle

Generation of massive entanglement, besides its interest for foundational
research of quantum theory, is of great importance for applications
in quantum information processing and precision measurements. Entanglement
is a valuable resource that can be used to enhance the performance
of quantum computation, the security of quantum communication, and
the precision of quantum measurements. For these applications, it
is desirable to get as many particles as possible into entangled states.
However, entanglement is typically fragile and many-particle entangled
states can be easily destroyed by decoherence due to inevitable coupling
to the environment. As an experimental record, so far fourteen qubits
carried by trapped ions have been successfully prepared into genuine
entangled states \cite{1}. Pushing up this number represents a challenging
goal in the experimental frontier.

The Bose Einstein condensate of ultracold atoms is in a pure quantum
mechanical state with strong collision interaction. In a spinor condensate
\cite{2,3,4}, the spin-dependent collision interaction can be used
to produce spin squeezing \cite{5,6}, which is an indicator of many-particle
entanglement \cite{7}. Spin squeezing has been demonstrated in condensates
in recent experiments through spin-dependent collision dynamics \cite{6,6b}.
A squeezed state is typically sensitive to noise and generation of
substantial squeezing requires accurate control of experimental systems,
which is typically challenging. In quantum information theory, the
Dicke states are known to be relatively robust to noise and they have
important applications for quantum metrology \cite{8} and implementation
of quantum information protocols \cite{9}. For instance, the three-particle
Dicke state, the so-called W state, has been proven to be the most
robust entangled state under the particle loss \cite{10}. Because
of their applications and nice noise properties, Dicke states represent
an important class of many-body states that are pursed in physical
implementation. For a few particles, Dicke states have been generated
in several experimental systems \cite{11} .

In this paper, we propose a robust method to generate massive entanglement
in the proximity of many-particle Dicke states through control of
adiabatic passage across a quantum phase transition in a spinor condensate.
Using conservation of the magnetic quantum number, we show that sweep
of the magnetic field across the polar-ferromagnetic phase transition
provides a simple method to generate many-body entanglement in this
mesoscopic system. The generated many-body entanglement can be characterized
through the entanglement depth, which measures how many particles
have been prepared into genuine entangled states \cite{7,13}. The
entanglement depth can be easily measured experimentally for this
system through a criterion introduced in Ref. \cite{12}. We quantitatively
analyze the entanglement production through the entanglement depth
and show that the scheme is robust under noise and experimental imperfection.
The scheme works for both the ferromagnetic (such as $^{87}Rb$ )
and the anti-ferromagnetic (such as $^{23}Na$ ) condensates. For
the anti-ferromagnetic case, we use adiabatic quantum phase transition
in the highest eigenstate of the Hamiltonian instead of its ground
state.

The system under consideration is a ferromagnetic (or anti-ferromagnetic)
spin-$1$ Bose Einstein condensate under an external magnetic field,
which has been realized with $^{87}Rb$ (or $^{23}Na$) atoms in an
optical trap\cite{4}. The Hamiltonian for the spin-$1$ condensate
can be divided into two parts $H=H_{0}+H_{i}$. The non-interacting
Hamiltonian $H_{0}$ and the interaction Hamiltonian $H_{i}$ take
respectively the following forms \cite{2,4}

\begin{eqnarray}
\hat{H_{0}} & = & {\displaystyle \sum_{m,n=0,\pm1}\int d\mathbf{r}\hat{\psi}_{m}^{\dagger}[-\frac{\hbar^{2}\nabla^{2}}{2M}}\nonumber \\
 & + & U(\mathbf{r})-p(f_{z})_{mn}+q(f_{z}^{2})_{mn}]\hat{\psi}_{n},\\
\hat{H}_{i} & = & \frac{1}{2}\int d\mathbf{r}[c_{0}:\hat{n}^{2}(\mathbf{r}):+c_{1}:\hat{F}^{2}(\mathbf{r}):].
\end{eqnarray}
where$\hat{\psi}_{m}(\mathbf{r})$ denote the bosonic filed operators
with the spin index $m=1,0,-1$, corresponding to annihilation of
an atom of mass $M$ in the Zeeman state $m$ on the hyperfine level
$F=1$. The atoms are trapped by the spin-independent optical potential$U(\mathbf{r})$.
The linear Zeeman coefficient $p=-g\mu_{B}B$, where $g$ is the Land\'e
g factor, $\mu_{B}$ is the Bohr magneton, and $B$ is the external
magnetic field. The quadratic Zeeman coefficient $q=\frac{(g\mu_{B}B)^{2}}{\Delta E_{hf}}$,
where $\Delta E_{hf}$ is the hyperfine energy splitting. The symbol
$f_{\mu}$ ($\mu=x,y,z$) denotes $\mu$-component of the spin-$1$
matrix, and $(f_{\mu})_{mn}$ is the corresponding $(m,n)$ matrix
element. The particle density operator $\hat{n}(\mathbf{r})$ and
the spin operator $\hat{F}(\mathbf{r})$ are defined respectively
by $\hat{n}(\mathbf{r})=\sum_{m=-1}^{1}\hat{\psi_{m}^{\dagger}}(\mathbf{r})\hat{\psi}_{m}(\mathbf{r})$
and $\hat{F}_{\mu}(\mathbf{r})=\sum_{m,n=-1}^{1}(f_{\mu})_{mn}\hat{\psi_{m}^{\dagger}}(\mathbf{r})\hat{\psi}_{n}(\mathbf{r})$.
The interaction coefficients $c_{0}=4\pi\hbar^{2}(a_{0}+2a_{2})/3M$,
$c_{1}=4\pi\hbar^{2}(a_{2}-a_{0})/3M$, where $a_{s}$ is the s-wave
scattering lengths for two colliding atoms with total spin $s$. We
have $c_{1}<0$ ($c_{1}>0$) for $^{87}Rb$ ($^{23}Na$), which corresponds
to ferromagnetic (anti-ferromagnetic) interaction, respectively.

For typical spinor condensates in experiments such as$^{87}Rb$ and
$^{23}Na$, we have $c_{0}\gg c_{1}$ , so the spin-independent interaction
dominates over the spin-dependent interaction. In this case, to describe
the ground state in a spin-independent optical trap $U(\mathbf{r})$,
it is good approximation to assume that different spin components
$\hat{\psi}_{m}(\mathbf{r})$ of the condensate take the same spatial
wave function $\phi(\mathbf{r})$. This is the well-known single mode
approximation \cite{3,4}, and under this approximation we have $\hat{\psi}_{m}\approx\hat{a}_{m}\phi(\mathbf{r}),(m=1,0,-1)$,
where $a_{m}$ is the annihilation operator for corresponding spin
mode and $\phi(\mathbf{r})$ is normalized as $\int d\mathbf{r}|\phi(\mathbf{r})|^{2}=1$.
We consider a spinor condensate with a fixed total particle number
$N$ as in experiments and neglect the terms in the Hamiltonian that
are constant under this condition. The spin-dependent part of the
Hamiltonian is then simplified to

\begin{equation}
H=c'_{1}\frac{\mathbf{L}^{2}}{N}+\sum_{m=-1}^{1}(qm^{2}-pm)a_{m}^{\dagger}a_{m}\label{eq:3}
\end{equation}
where we have introduced the spin-$1$ angular momentum operator $\mathbf{L_{\mu}}=\sum_{m,n}a_{m}^{\dagger}(f_{\mu})_{mn}a_{n}$
and defined $c'_{1}=c_{1}N\int d\mathbf{r}|\phi(\mathbf{r})|^{4}/2$.
The linear Zeeman term $\sum_{m=-1}^{1}pma_{m}^{\dagger}a_{m}=pL_{z}$
typical dominates in the Hamiltonian $H$, however, this term commutes
with all the other terms in the Hamiltonian, and if we start with
an initial state that is an eigenstate of $L_{z}$, the linear Zeeman
term has no effect and thus can be neglected. In this paper, we consider
an initial state with all the atoms prepared to the level $\left|F=1,m=0\right\rangle $
through optical pumping, which is an eigenstate of $L_{z}$. The system
remains in this eigenstate with magnetization $L_{z}=0$, and the
effective spin Hamiltonian becomes
\begin{equation}
H=c'_{1}\frac{\mathbf{L}^{2}}{N}-qa_{0}^{\dagger}a_{0}.\label{eq:4}
\end{equation}
The ratio $q/c'_{1}$ is the only tunable parameter in this Hamiltonian,
and depending on its value, the Hamiltonian has different phases resulting
from competition between the quadratic Zeeman effect and the spin-dependent
collision interaction.

We first consider the ferromagnetic case with $c'_{1}<0$. For the
initial state, we tune up the magnetic field to make the quadratic
Zeeman coefficient $q\gg\left|c'_{1}\right|$. In this limit, the
second term $-qa_{0}^{\dagger}a_{0}$ dominates in the Hamiltonian
$H$. The ground state of the Hamiltonian is given by an eigenstate
of $a_{0}^{\dagger}a_{0}$ with the maximum eigenvalue $N$. This
ground state can be prepared by putting all the atoms to the Zeeman
level $\left|F=1,m=0\right\rangle $through optical pumping. Then
we slowly ramp down the magnetic field to zero. From the adiabatic
theorem, the system remains in the ground state of the Hamiltonian
$H$ and the final state is the lowest-energy state of $H_{F}=c'_{1}\mathbf{L}^{2}/N$,
which is the Dicke state $\left|L=N,L_{z}=0\right\rangle $ that maximizes
$\mathbf{L}^{2}$ with the eigenvalue $L(L+1)$. The Dicke state $\left|L=N,L_{z}=0\right\rangle $
is a massively entangled state of all the particles.

The above simple argument illustrates the possibility to generate
massive entanglement through an adiabatic passage. To turn this possibility
into reality, however, there are several key issues we need to analyze
carefully. First, we need to know what is the requirement of the sweeping
speed of the parameter $q$ to maintain an adiabatic passage. In particular,
this adiabatic passage goes through a quantum phase transition where
the energy gap approaches zero in the thermodynamical limit. So the
evolution cannot be fully adiabatic for a large system. It is important
to know how the energy gap scales with the particle number $N$ for
this mesoscopic system. Second, due to the non-adiabatic correction
and other inevitable noise in a real experimental system, the final
state is never a pure state and quite different from its ideal form
$\left|L=N,L_{z}=0\right\rangle $. For a many-body system with a
large number of particles, the state fidelity is always close to zero
with presence of just small noise. So we need to analyze whether we
can still generate and confirm genuine many-particle entanglement
under realistic experimental conditions.

To analyze the entanglement behavior, first we quantitatively calculate
the phase transitions during this adiabatic passage and analyze how
the energy gap scales with the particle number $N$. The mean-field
phase diagram for the Hamiltonian (3) is well known \cite{4}. However,
in typical mean field calculations one fixes the parameters $p,q$
to obtain the ground state of the Hamiltonian (3), and this ground
state in general has varying magnetization $\left\langle L_{z}\right\rangle $.
For our proposed adiabatic passage, we should fix $L_{z}=0$ and find
the ground state of the Hamiltonian (4) instead of (3) as the linear
Zeeman term is irrelevant. We perform exact numerical many-body calculation
in the Hilbert space with $L_{z}=0$ to find the ground state of the
Hamiltonian (4) and draw the condensate fraction in the the Zeeman
level $\left|F=1,m=0\right\rangle $, $N_{0}/N$ with $N_{0}\equiv\left\langle a_{0}^{\dagger}a_{0}\right\rangle $,
in Fig.1 as we ramp down the parameter $q$. Control of the magnetic
field can only sweep the parameter $q$ from the positive side to
zero. Further sweep of $q$ to the negative side can be obtained through
ac-Stack effect induced by a microwave field coupling the hyperfine
levels $\left|F=1\right\rangle $ and $\left|F=2\right\rangle $,
as demonstrated in experiments \cite{14}. The curve in Fig. 1 shows
two second-order phase transitions at the positions $q/\left|c'_{1}\right|=\pm4$
, where the condensate fraction $N_{0}/N$ drop first from $1$ to
a positive number $r$ $\left(0<r<1\right)$ and then to $0$. The
transition point at $q/\left|c'_{1}\right|=4$ agrees with the mean
field prediction, however, there is a significant discrepancy for
the transition at $q/\left|c'_{1}\right|=-4$ . Mean field calculation
under a fixed parameter $p=0$ predicts a transition at $q/\left|c'_{1}\right|=0$,
where the magnetization $\left\langle L_{z}\right\rangle $ abruptly
changes \cite{4}. For the adiabatic passage considered here, due
to the conservation of $L_{z}$ the transition at $q/\left|c'_{1}\right|=0$
is postponed to the point $q/\left|c'_{1}\right|=-4$. 
\begin{figure}
\includegraphics[width=7cm,height=3cm]{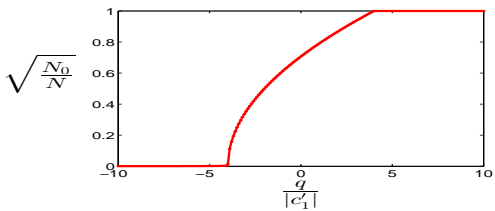}\caption{The order parameter $\sqrt{\left<N_{0}/N\right>}$ shown as a function
of the quadratic Zeeman coefficient $q$ in the unit of $\left|c'_{1}\right|$
for the total atom number $N=10^{5}$. Two second-order phase transitions
take place at $q/\left|c'_{1}\right|=\pm4$. }
\end{figure}

Besides prediction of the phase transition points, the exact many-body
calculation can show evolution of entanglement for the ground state
and scaling of the energy gap with the particle number $N$ at the
phase transition points. The scaling of the energy gap is important
as it shows the relevant time scale to maintain the adiabatic passage.
In Fig. 2(a), we show the energy gap $\Delta$ (defined as the energy
difference between the ground state and the first excited state) in
the unit of $\left|c'_{1}\right|$ as a function of $q/\left|c'_{1}\right|$
for $N=10^{4}$ particles. The gap attains the minimum at the phase
transition points and is symmetric with respect to the transitions
at $q/\left|c'_{1}\right|=\pm4$ . In Fig. 2(b), we show how the energy
gap at the phase transition point scales with the particle number
$N$. In the $\log-\log$ plot, the points are on a line, which can
be well fit with the polynomial scaling $\Delta=7.4N^{-1/3}$. The
energy gap decreases slowly with increase of the particle number $N$,
which suggests it is possible to maintain an adiabatic passage for
typical experimental systems with $N\sim10^{5}$.
\begin{figure}
\includegraphics[width=8.5cm]{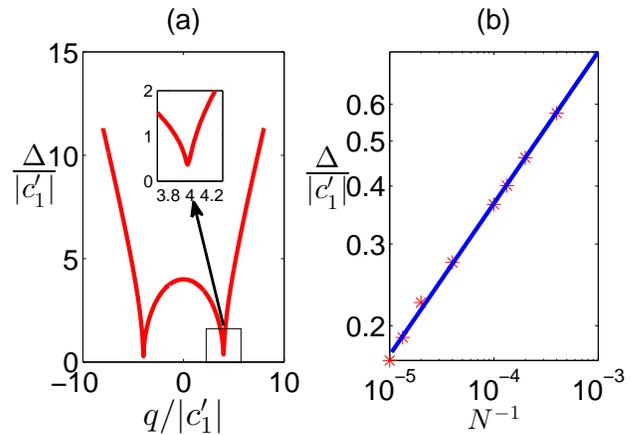}\caption{(a) The energy gap $\Delta$ in the unit of $\left|c'_{1}\right|$
shown as a function of $q/\left|c'_{1}\right|$ with the total particle
number $N=10^{4}$. (b) The stars show the scaling of the energy gap
$\Delta/\left|c'_{1}\right|$ at the phase transition point with the
particle number $N$ in the log-log plot. The solid line is a linear
fit to the data points with $\Delta=7.4N^{-1/3}$. }
\end{figure}

With this understanding, we now turn to our main task to characterize
entanglement generation with this adiabatic passage. For this purpose,
we need to have a quantity to measure entanglement in the proximity
of the Dicke state and this measure should be accessible to experimental
detection. Due to non-adiabatic corrections and inevitable noise in
real experiments, we cannot assume that the system is in a pure state
and the entanglement measure should work for any mixed states. Many-body
entanglement can be characterized in different ways, and a convenient
measure is the so-called entanglement depth which measures how many
particles in an $N$-particle system have been prepared into genuine
entangled states given an arbitrary mixed state of the system \cite{7,12,13}.
A quantity to measure the entanglement depth for $N$ spin-$1/2$
particles has been provided in Ref. \cite{12} based on measurements
of the collective spin operators. It is straightforward to generalize
this quantity to the case of $N$ spin-$1$ particles. For $N$ spin-$1$
particles, the collective spin operator is defined by $\boldsymbol{L}=\sum_{i=1}^{N}l_{i}$,
where $l_{i}$ denotes the individual spin-$1$operator. In terms
of the bosonic mode operators, the collective spin operator has the
standard decomposition $\mathbf{L_{\mu}}=\sum_{m,n}a_{m}^{\dagger}(f_{\mu})_{mn}a_{n}$
$\left(\mu=x,y,z;\: m,n=0,\pm1\right)$. To characterize entanglement
in the proximity of the Dicke state $\left|L=N,L_{z}=0\right\rangle $,
we measure the quantity
\begin{equation}
\xi=\frac{\left<L_{x}^{2}\right>+\left<L_{y}^{2}\right>}{N(1+4\left<(\Delta L_{z}^{2})\right>)}\label{eq:5}
\end{equation}
If $\xi>m$, from the arguments that lead to theorem $1$ of Ref.
\cite{12} we conclude that the system has at least genuine $m$-particle
entanglement (i.e., the entanglement depth is bounded by $m$ from
below). For the ideal Dicke state $\left|L=N,L_{z}=0\right\rangle $,
one can easily verify that $\xi=N+1>N$, so all the $N$ particles
are in a genuine entangled state. The final state of real experiments
is in general a complicated mixed state which is impossible to be
read out for many-particle systems. The power of the measure in Eq.
(5) is that it gives an experimentally convenient way to bound the
entanglement depth in this case through simple detection of the collective
spin operators even through the system state remains unknown.

Now we show how the entanglement measure defined in Eq. (5) evolves
when we adiabatically sweep the parameter $q$ in the Hamiltonian
(4). We ramp down the parameter $q$ linearly from $q=6\left|c'_{1}\right|$
to $0$ with a constant speed, starting from the initial product state
with all the particles in the level $\left|F=1,m=0\right\rangle $.
The entanglement depth $\xi$ (in the unit of $N$) of the final state
is shown in Fig. 3(a) and 3(b) as a function of the sweeping speed
$v$ (in the unit of $\left|c'_{1}\right|^2$ by taking $\hbar=1$)
for $N=10^{3}$ and $N=10^{4}$, respectively. We can see that the
entanglement depth increases abruptly from a few to the order of $N$
when the speed $v$ decreases below $\left|c'_{1}\right|^2$. In the
same figure, we also show the excitation probability of the final
state (the probability to be not in the ground state). For a small
number of particles, the excitation probability typically correlates
with the entanglement depth, and they jump roughly around the same
value of the sweeping speed. However, for a large number of particles
(e.g., $N\geq10^{4}$), we can have the entanglement depth of the
order of $N$ while the excitation probability is near the unity as
shown in Fig. 3(b). This indicates that the entanglement in the proximity
of the Dicke state is quite robust. Even when the sweep is not fully
adiabatic and most of the atoms are excited to the low-lying excited
states (meaning that the state fidelity decrease to almost zero),
we can still have the entanglement depth close to $N$ (meaning all
the particles are still genuinely entangled).

\begin{figure}
\includegraphics[width=8.5cm]{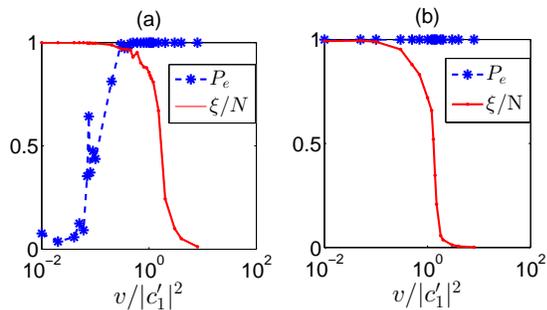}\caption{The normalized entanglement depth $\xi/N$ (solid lines) and the excitation
probability $P_{e}$ (star points) for the final state shown as functions
of the sweeping speed $v$ (in the unit of $\left|c'_{1} \right|^2$)
for the number of particles $N=10^{3}$ (a) and $N=10^{4}$ (b). The
parameter $q$ in the Hamiltonian (4) is ramped down linearly from
$q=6\left|c'_{1}\right|$ to $0$ at a constant speed $v$, starting
from the initial product state with all the particles in the level
$\left|m=0\right\rangle $. }
\end{figure}

\begin{figure}
\includegraphics[width=7cm,height=3cm]{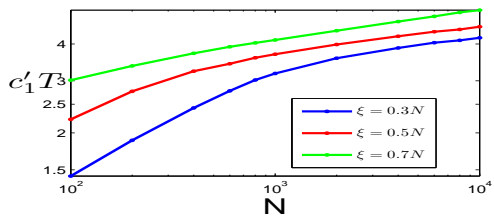}\caption{Scaling of the required sweeping time $T$ (in the unit of $1/\left|c'_{1}\right|$)
with the particle number $N$ as we fix the entanglement depth of
the final state to be$0.3N$ (bottom curve), $0.5N$ (middle curve),
and$0.7N$ (top curve), respectively. }
\end{figure}

As the energy gap $\Delta$ at the phase transition point decreases
with the atom number $N$, one expects that the required sweeping
time $T$ to get substantial entanglement increases with $N$. However,
this increase is very slow. First, $\Delta$ increases slowly with
$N$ by the scaling $\Delta\propto N^{-1/3}$ as shown in Fig. 2(b).
Second, for a large $N$ even when $\Delta T<1$ and a significant
fraction of the atoms get excited during the sweep, we can still observe
substantial entanglement as the entanglement depth of the low-lying
excited states is still high as shown in Fig. 3(b). To see the the
quantitative relation between the required sweeping time $T$ and
the particle number $N$, we fix the entanglement depth of the final
state to be a significant number (e.g., with $\xi=0.3N,\:0.5N,$ or
$0.7N$) and draw in Fig. 4 the scaling of $T$ (in the unit of $1/\left|c'_{1}\right|$)
as a function of $N$. When $N\geq10^{3}$, the curve of $\left|c'_{1}\right|T$
is almost flat, increasing by a modest $20\%$ when the atom number
grows by an order of magnitude.

All the calculations above are done for the ferromagnetic case with
$c'_{1}<0$ by assuming an adiabatic sweep of the Hamiltonian (4)
in its ground state. For the anti-ferromagnetic case with $c'_{1}$>0
(such as $^{23}Na$), we can perform an adiabatic sweep along the
ground state of the Hamiltonian $-H$ (or the highest eigenstate of
the Hamiltonian $H$ in Eq. (4)). Then, all the calculations above
equally apply to the anti-ferromagnetic case. The only difference
is that initially the parameter $q$ needs to be set to the negative
side with $q=-6\left|c'_{1}\right|$ when the atoms are prepared into
the level $\left|m=0\right\rangle $. As mentioned before, $q$ can
be switched to both the positive and the negative sides, through ac-Stack
shift from a $\pi$-polarized microwave field that couples the hyperfine
levels $\left|F=1\right\rangle $ and $\left|F=2\right\rangle $ \cite{14}.
An advantage of using $^{23}Na$ instead of $^{87}Rb$ is that it
is has a larger spin-dependent collision rate $\left|c'_{1}\right|$
and thus allows a faster sweep of the parameter $q$. If we take the
peak condensate density about $10^{14}cm^{-3}$, $c'_{1}/\hbar$ is
estimated to be about $-2\pi\times7Hz$ for $^{87}Rb$ atoms and $2\pi\times50Hz$
for $^{23}Na$ atoms.

Finally, we briefly discuss how the noise influence entanglement generation
in this scheme. First, in the proximity of the Dicke
state the entanglement depth measured through Eq. (5) is very robust
to the dephasing noise (dephasing between the Zeeman levels caused
by, e.g., a small fluctuating magnetic field). As shown in Ref. \cite{12},
even with a dephasing error rate about $50\%$ for each individual
atom, the entanglement depth $\xi$ remains about $N/2$, which is
still large. The entanglement depth is more sensitive to the bit-flip
error that increases $\left<\Delta L_{z}^{2}\right>$in Eq. (5), which
can be caused by imperfect preparation of the initial state, atom
loss during the adiabatic sweep, or imperfection in the final measurement
of the collective spin operators. The detection error can be corrected
through simple data processing using the method proposed in Ref. \cite{15}
as long as its error rate has been calibrated. The initial state $\left|F=1,m=0\right\rangle $
can be prepared efficiently through optical pumping and remaining
atoms in the $\left|F=1,m=\pm1\right\rangle $ levels can be blown
away through microwave coupling to the $\left|F=2\right\rangle $
levels that are unstable under atomic collisions. The atomic loss
should be small as the sweeping time $T$ is assumed to be much shorter
compared with the life time of the condensate. Only loss of atoms
in the components $\left|F=1,m=\pm1\right\rangle $ can increase the
fluctuation $\left<\Delta L_{z}^{2}\right>.$ Assume the loss rate
is $p$ during the sweep, the resultant $\left<\Delta L_{z}^{2}\right>$
is estimated by $\left<\Delta L_{z}^{2}\right>\sim Np(1-p)/6$. For
a large number of atoms with $Np\gg1$, the entanglement depth in
Eq. (5) is then estimated by $\xi\sim3/\left(2p\right)$ . If we take
$p$ about $1\%$, it is possible to prepare a remarkable number
of hundreds of atoms into genuine entangled states.

We thank Y.-M. Liu for discussions. This
work was supported by the NBRPC (973 Program) 2011CBA00300 (2011CBA00302), 
the DARPA OLE program, the IARPA MUSIQC program, the ARO and the AFOSR MURI program.

\end{document}